\documentclass[twocolumn,preprintnumbers,amsmath,amssymb]{revtex4-2}

\usepackage{graphicx}  
\usepackage{sidecap}
\begin{document}
\title{S-shaped current-voltage characteristics of n$^+$-i-n-n$^+$ graphene  field-effect transistors due to the Coulomb drag 
 of quasi-equilibrium electrons by ballistic electrons
}
\author{V. Ryzhii$^{1,2}$,  M. Ryzhii$^3$, V. Mitin$^4$, M. S. Shur$^5$, and T.~Otsuji$^1$}
\address{
$^1$Research Institute of Electrical Communication,~Tohoku~ University,~Sendai 980-8577, Japan\\
$^2$Institute of Ultra High Frequency Semiconductor Electronics of RAS,
 Moscow 117105, Russia\\
 $^3$Department of Computer Science and Engineering, University of Aizu, Aizu-Wakamatsu 965-8580, Japan\\
$^4$Department of Electrical Engineering, University at Buffalo, SUNY, Buffalo, New York 14260 USA\\
$^5$Department of Electrical,~Computer,~and~Systems~Engineering, Rensselaer Polytechnic Institute, Troy, New York 12180, USA
}
 \begin{abstract} 
We demonstrate that  the injection of the ballistic electrons into the two-dimensional electron plasma in lateral  n$^+$-i-n-n$^+$ graphene field-effect transistors (G-FET)  might lead to a substantial Coulomb  drag of the quasi-equilibrium electrons due the violation of the Galilean and Lorentz invariance in the  systems with a linear electron dispersion. This effect can result in the  S-shaped current-voltage characteristics (IVs).
The resulting negative differential conductivity enables the hysteresis effects and current filamentation
that can be used  for the implementation of voltage switching devices.
   Due to a strong nonlinearity of the IVs, the G-FETs can be used for an effective frequency multiplication and detection of terahertz radiation.
\end{abstract} 

\maketitle

\newpage
\section{Introduction}

The lateral transport of electrons and holes in the graphene layer (GL) heterostructures 
could enable 
 the detection, amplification, and generation of terahertz radiation~\cite{1,2} and other numerous applications (see, for example,~\cite{3}). 
In this paper, we analyze the electron transport in the lateral  n$^+$-i-n-n$^+$ graphene  field-effect transistors (G-FETs) with the n$^+$ source and drain contacts and the gated n-region.
Figure~1 shows the G-FET
structure and the band diagrams at different source-drain  voltage:  $V < \hbar\omega_0/e$  and $V > \hbar\omega_0/e$, where $\hbar\omega_0 \simeq 200$~meV is the optical phonon energy in graphene and $e$ is the electron charge.
  The  n-region  is formed by the  
   positive gate bias $V_g$.
Similar lateral heterostructure GL devices including those based on more complex lateral periodical cascade devices were reported previously~\cite{1,2,3,4,5,6,7}. 
One of the remarkable advantages of the GLs is the possibility of very high directed velocities of the electron (hole) ensembles close to the characteristic velocity $v_W \simeq 10^8$~cm/s~\cite{6,7} providing the collision-less, i.e., ballistic electron (BE) motion 
~\cite{8,9} in relatively long channels.
As demonstrated experimentally,  in the graphene encapsulated in hexagonal boron nitride 
the ballistic transport is realized in the samples with the length of a few $\mu$m at room temperature~\cite{10}
and of 28 $\mu$m at decreased temperatures~\cite{11}.

As was predicted decades ago~\cite{9}, the BE motion interrupted by the emission of the optical phonons
can enable the self-excitation of the current oscillations leading to the radiation emission~\cite{12,13,14,15,16}.

\begin{figure}[t]
\centering
\includegraphics[width=7.5cm]{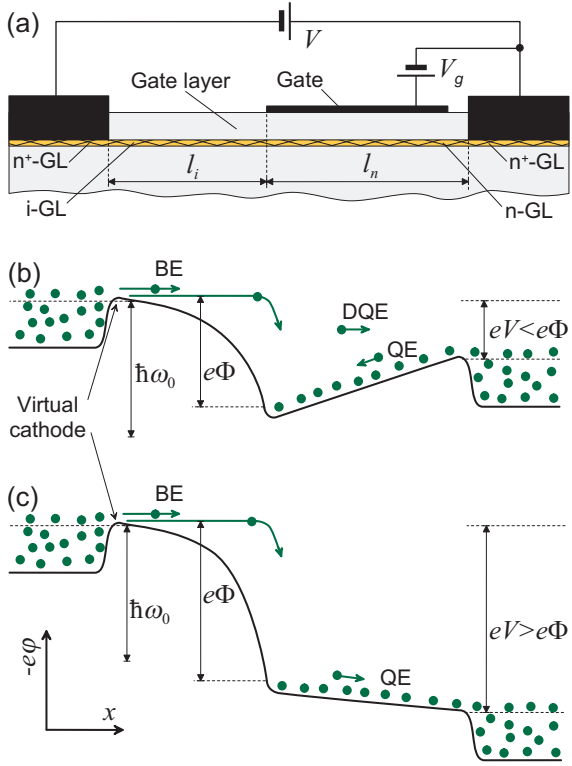}
\caption{(a) Schematic views of the lateral  n$^+$-i-n-n$^+$  G-FETs with electrically induced- n-region and their band diagrams corresponding to (b) $T < eV < \hbar\omega_0$ (intermediate current densities at a pronounced drag)
and   (c) to $eV > \hbar\omega_0$ monotonic (elevated current densities) potential distributions. The BEs are injected via the virtual cathode. Arrows correspond to the BEs injected from the source, the  DQEs, i.e., the QEs  dragged by the injected BEs, and the QEs injected from the drain.}
\label{F1}
\end{figure}

Considering the lateral forward-biased n$^+$-i-n-n$^+$ G-FET  with the sufficiently perfect GL, we assume that 
the transport of the injected electrons from the emitter n$^+$ region into  the i-region ($-l_i < x < 0$, where $l_i$ is  the i-region length) is  ballistic. This implies that the BE transit time in the i-region is much shorter that the characteristic times of their scattering on the impurities and the acoustic phonons, 
$\tau_{imp}$ and $\tau_{ac}$.  The impurity and acoustic phonon scattering of the BEs injected into the n-region ($0 < x < l_n$, where $l_n$ stands for the n-region length) is also insignificant. Thus, $l_i, l_n \ll v_W\tau_{imp}, v_W\tau_{ac}$.  
We demonstrate that the Coulomb collisions of the BEs, injected into  the n-region, 
with the thermalized quasi-equilibrium electrons (QEs) 
 can lead to 
the "conversion" of  a fraction of  these electrons  into
 the dragged equilibrium electrons (DQEs)  moving toward the n$^+$ collector drain region (analogous to the mutual electron-hole drag). Such a Coulomb drag in GLs, i.e., in  the electron systems with the linear energy
spectrum can be fairly effective.
 The GL electron system 
is neither a Galilean  nor a truly Lorentz-invariant system~\cite{17,18,19}. 
The Coulomb drag in question is fundamentally similar to the drag between spatially separated  
 standard~\cite{20,21} and graphene-based~\cite{22,23,24,25,26,27,28,29}  
two-dimensional electron-hole
systems. This  effect was 
extensively studied in graphene both theoretically and experimentally (see, for example,~\cite{22,23,27}).
An essential distinction of the ballistic-equilibrium drag is the current non-conservation (and possible multiplication) due to electron-electron collisions.  
The  Coulomb electron drag in the G-FETs under consideration with the current multiplication might 
 pronouncedly 
affect the device characteristics resulting in the  S-type current-voltage characteristics (IVs).
The latter can lead to the hysteresis phenomena and the  instability of the uniform current flow (the current filamentation).

Similar phenomenon can take place in the reverse-biased  
p$^+$-p-i-n-n$^+$ devices~\cite{6,7} due to the interband tunneling generation
~\cite{23,24} 
of the electron-hole pairs in the i-region. 

The physics behind is the Coulomb drag by the BEs. BEs collisions with   QEs results in the  latter contributing to the current. As a result, the current voltage characteristic is nonlinear even at low applied voltages, since the voltage increase results in a higher level of the BE injection. At a certain threshold voltage, this nonlinearity might lead to the infinite differential conductance. At the threshold voltage, the switching occurs into another stable branch of the current-voltage characteristic with the dominant contribution of the BEs. As a consequence, there are two stable branches of the IV: (a) the low current branch with a relatively few injected BEs and (b) the high current branch with the dominant BE transport. The switching occurs at the threshold voltage and the value of current after switching depends on the load line (i.e. on the load resistance). During the switching, the current traverses the unstable branch with the negative differential conductance. Depending on the load resistance and on the applied voltage, the final state might correspond to the current filamentation when the device cross section is divided into  regions corresponding to the low current and high current branches, respectively

The paper is organized as follows. In Sec.~II, we find the potential distribution in the G-FET injection region (i-region) and derive
the injected current density as a function of the potential drop across this region. Section~III deals with the analysis of the Coulomb drag of the QEs by the BEs. In this section,  the net current density in the n-region is expressed as a sum of the BEs, DQEs, and QEs. Using the results of Sec.~III,  we derive the IVs in Sec.~IV and demonstrate that they can be both   monotonic and S-shaped. In Sec.~V, we consider the possibility of the current switching in the G-FETs enabled by the S-shape of their IVs. 
Section~VI deals with  the instability of  uniform current spatial distributions at the fixed terminal current,
which, as indicated, can lead to the formation of the stationary and pulsing current filaments.
Section VII is devoted to the comments associated with the device model. In Conclusions (Sec.~VIII) we summarize the main results of the paper. Some,  mainly intermediate mathematical  results, are given in  Appendix A and Appendix B.

\section{Potential distribution and injected current}

The injection current density $j_i$
 is determined by the voltage drop $\Phi = \varphi|_{x=0}$ across the GL  i-region and by the space charge in this region (the space-charge-limited electron injection~\cite{30}). The potential $\Phi$ is determined by the potential spatial distribution across the entire G-FET structure corresponding to the boundary conditions $\varphi_{x = -l_i} =0$ and $\varphi_{x =  l_n} = V$,
 where  $l_i$ and $l_n$ are the lengths of the i- and n-regions.
For the lateral G-FET structure with the blade-like regions near the i-region edges and for  the injected BE density  $
\Sigma_i = j_i/e v_W$ (where $v_W \simeq 10^8$~cm/s is the characteristic electron velocity in GLs and $e=|e|$ is the electron charge)
 the potential distribution across the i-region satisfying the conditions $\varphi_{x = -l_i} =0$ and $\varphi_{x = 0} = \Phi$,  and  $j_i$ versus $\Phi$ relation  can be found as follows (compare with, for example,~\cite{30,31,32,33}):

\begin{eqnarray}\label{eq1}
\varphi  =\frac{2\Phi}{\pi}\cos^{-1}\biggl(-\frac{x}{l_i}\biggr) - \frac{2e\Sigma_i}{\kappa} x\ln\biggl[\frac{l_i -\sqrt{l_i^2 - x^2}}{l_i + \sqrt{l_i^2 - x^2}}\biggr],
\end{eqnarray}

\begin{eqnarray}\label{eq2}
E = \biggl[\frac{4e\Sigma_i l_i
}{\kappa}
 - \frac{2\Phi}{\pi}\biggr] \frac{1}{\sqrt{l_i^2-x^2}}\nonumber\\
+ \frac{2e\Sigma_i }{\kappa} \ln\biggl[\frac{l_i -\sqrt{l_i^2 - x^2}}{l_i + \sqrt{l_i^2 - x^2}}\biggr].
\end{eqnarray}
Here $\kappa$ is the dielectric constant of the material surrounding the GL.
To consider the regime of electron injection limited by the electron space charge near the
source n-i-junction, we set the electric field at a point $x= -l_i +0$ (the "virtual cathode"~\cite{30}, see Fig.~1)
to be equal to zero. Accounting for Eq.~(2), this condition yields

\begin{equation}\label{eq3}
j_i =v_W\biggl(\frac{\kappa\Phi}{2\pi\,l_i}\biggr),\qquad 
\Sigma_i = 
 \frac{\kappa\Phi}{2\pi\,el_i}.
\end{equation}

Equations~(1) - (3) are valid when $e\Phi, eV > T$, where $T$ is the temperature in the energy units.
Equation~(3) is in line with the well known result obtained for the devices with blade-like injection contacts
(but for the carrier transport with the saturation velocity $v_S \ll v_W$). However, Eq.~(3) yields different voltage dependence from those found for different bulk contacts~\cite{4}.
 
 Hence, according to Eqs.~(1) and (2), we obtain

\begin{eqnarray}\label{eq4}
\varphi = \frac{2\Phi}{\pi}\biggl[\cos^{-1}\biggl(-\frac{x}{l_i}\biggr)
-\frac{x}{2l_i}\ln\biggl(\frac{l_i -\sqrt{l_i^2 - x^2}}{l_i + \sqrt{l_i^2 - x^2}}\biggr)
\biggr],
\end{eqnarray}

\begin{eqnarray}\label{eq5}
E = 
 \frac{\Phi}{l_i} \ln\biggl[\frac{l_i -\sqrt{l_i^2 - x^2}}{l_i + \sqrt{l_i^2 - x^2}}\biggr].
\end{eqnarray}
Thus, $E|_{x = -l_i} = 0$ and $E|_{x = 0} \simeq (\Phi/l_i)\ln(d/2l_i)^2$ 
$(|E|_{x = 0}| \gg\Phi_i/l_i$). Here, 
 $d$ is  the thickness of the gate layer.

\section{Coulomb electron drag}

The BEs  injected into the  n-region have the energy $\varepsilon_i = e\Phi$ and the momentum $p_i \simeq e\Phi/v_W$. 
The collisions between the injected BEs and QEs in the  n-region
result in the transfer of a part of the ballistic electron momentum to the QEs.
Due to the linearity of the electron spectrum in GLs, the injected and the BEs scattered by the QEs with small energies  preserve the direction of their movement (in the direction $x$ from the emitter to the collector), while their directed  momentum changes
from $p_i$ to $p_i^{\prime} < p_i$.
Despite the lost  portion of the momentum, the BE  continues its motion toward the collector with the same velocity $v_W$.
Due to the collision with the BE, the QE receives the momentum $p_s = p_i - p_i^{\prime}$. According to the energy and momentum conservation laws for the linear electron energy dispersion relation, the QEs also move with the velocity $v_W$ in the $x$-direction, so that, in contrast to both bulk and conventional two-dimensional   semiconductor systems, the momentum conservation at the electron-electron collisions does not lead  to the velocity conservation~\cite{17}.
In other words, a portion of the QEs becomes excited with the average directed momentum and velocity upon collisions with the injected BEs.

Thus, the electron collisions between the injected BEs and the QEs convert 
QEs into BEs
 doubling of the net current carried by the original and "secondary" BEs. 
Hence, the BE current density $j_{n}^{(BE)}$ (associated with   the original BEs, which came from the i-region, and the  QEs dragged by the BEs, to which we refer to as the DQEs)  in the  n-region ($0 < x <  l_n$) can exceed the BE current density
$j_i(\Phi)$  in
the i-region. This we interpret as the amplified   QE drag by the injected BEs.

The spatial variation of the BE current density  $j_n^{(BE)}$ across the n-region 
due to the  BEs scattering on the QEs  and optical phonons 
 is determined by

\begin{equation}\label{eq6}
\frac{dj_n^{(BE)}}{dx} = -\frac{K}{l_n}j_n^{(BE)}
\end{equation}
with
$K=   K_{ee}+ K_{ac} + K_{op},$
where  
\begin{equation}\label{eq7}
K_{ee}= \frac{l_n}{v_W\tau_{ee}}, \qquad K_{ac}= \frac{l_n}{v_W\tau_{ac}},
\end{equation}

\begin{eqnarray}\label{eq8}
K_{op} = \frac{l_n}{v_W\tau_{op}}
\frac{(e\Phi - \hbar\omega_0 +\mu)}{\hbar\omega_0}\Theta(e\Phi - \hbar\omega_0)\nonumber\\
={\overline K}_{op}\frac{(e\Phi - \hbar\omega_0 +\mu)}{\hbar\omega_0}\Theta(e\Phi - \hbar\omega_0).
\end{eqnarray}
Here
$\tau_{ee}$, $\tau_{ac}$,
   and $\tau_{op} = (\rho\hbar\,v_W^2/D_0)$~\cite{33,34,35,36} are the characteristic times of the electron-electron (BEs on QEs)  
   scattering,  and the BE scattering on acoustic and optical phonons, 
$\rho$ and $D_0$ are the GL density and the optical deformation potential, respectively,
 and $\Theta(e\Phi - \hbar\omega_0)$ is the unity step function reflecting the threshold  character of the optical phonon emission. 
 To account for the temperature and electron spectrum  smearing of the optical phonon emission threshold,
we set $\Theta(z) = [1+ \exp(-2z/T)]^{-1}$ with $T$ being the QE temperature. 
 The linear factor $\propto (e\Phi - \hbar\omega_0 + \mu)$ in the expression for $K_{op}$ is associated with the linearity of the GL density of states near the Dirac point. The Fermi electron energy in the gated  n-region,  $\mu$, appears in the latter function argument to account for  the optical phonon emission with the electron transitions to the states above the Fermi level.  
For simplicity we neglect  the BE scattering on
impurities not only in the i-region, but in the n-region as well because
in the G-FETs under consideration  with the gated n-region the electrons   
are primarily induced by the gate voltage (not by ionized impurities).
The quantity $K_{ee}$ markedly exceeds $K_{ac}$. At the electron densities $\Sigma_n \simeq 1\times(10^{12} - 10^{13})$~cm$^{-2}$ and room  temperature $T$ 
one can set
for the  energy of the BEs injected into the n-region $\varepsilon\sim \hbar\omega_0 $   $\tau_{ee}^{-1} \simeq (10 - 50)$~ps$^{-1}$,  $\tau_{ac}^{-1} \simeq 0.5$~ps$^{-1}$, and $\tau_{op}^{-1} \simeq (1-2)$~ps$^{-1}$~\cite{17,33,34,35,36,37,38}.
 If $l_n = (0.5 -1.0)~\mu$m, we find $K_{ee} \simeq 5 - 50$,
$K_{ac} \simeq 0.25 - 0.5 $, and ${\overline K}_{op} \simeq 0.5-2$. At lower temperatures, $K_{ac}$ becomes even smaller.
For brevity we do not distinguish 
the different  intra-valley and inter-valley optical phonon modes using  for their energies the common value $\hbar\omega_0 \simeq 200$~meV and accounting for the contribution of both modes by choosing the proper scattering time $\tau_{op}$.

Since $j_n^{(BE)}|_{x=0} = j_i$, where $j_i$ is given by Eq.~(3),
as follows from Eq.~(6), one obtains
for the density of the BE current injected into the n$^+$-contact at $x = l_n$ 

\begin{equation}\label{eq9}
j_n^{(BE)} = j_i e^{-K}.
\end{equation}
The BEs colliding with the QEs in the n-region transfer to the latter the average (per one QE) momentum equal to

\begin{equation}\label{eq10}
<p_x> = 
\frac{j_i\Phi}{v_W^2\Sigma_n}e^{ -K_{ac} - K_{op}}(1 - e^{- K_{ee}}),
\end{equation}
where $\Sigma_n$ is the QE density. We have disregarded a weak spatial  nonuniformity of the electron density in this region $\Sigma_n= \Sigma_d + \Sigma_g \simeq \Sigma_g$, where
$\Sigma_d$ the density of the ionized donors and 
$\Sigma_g= [\kappa(V_g - \varphi)]/(4\pi\,ed) \simeq \kappa\,V_g/(4\pi\,ed) \simeq const$ is the electron density induced by the gate voltage (the effect of the quantum capacitance is disregarded for simplicity as well).

The QE drag resulting in  the  QEs direct momentum induces the QE
current, so that the density of the net current $j_n^{>}$ (associated with the injection of the BEs), into the collector n$^+$-region can be presented as

\begin{equation}\label{eq11}
j_n^{>}= j_ie^{-K} + e\Sigma_n<v_x>.
\end{equation}
Here $v_x$ is the QE average velocity caused by the QE drag.
It is related to $<p_x>$ [see the Appendix, Eqs.~(A3) and (A5)] as

\begin{equation}\label{eq12}
<v_x> =\frac{<p_x> v_W^2}{T\xi}e^{-K_{ac}}.
\end{equation}
Here  $\xi = \xi(\mu/T)$ is a coefficient determined by
the QE statistics, where $\mu$ is the QE Fermi energy 
(see, Appendix A).

\begin{table*}[t!]
\caption{\label{table}Current-voltage characteristics peculiar points}
\centering
\renewcommand{\arraystretch}{3}
\begin{tabular}{l|c|c|c} 
\hline\hline  
Voltage, $V$  \qquad &  \qquad $0$  \qquad  \qquad & $V_{min}=\displaystyle V_0\frac{(1+\eta -b)}{\eta}$ & $V_{max}= \displaystyle V_0\frac{(1+\eta)^2}{4b\eta}$ \\
\hline
Current density, $j$  \qquad &  \qquad $0$  \qquad  \qquad &  \qquad $j_{min} = \displaystyle j_0\frac{(1+\eta -b)}{b}, \qquad j_{max}^{\infty}\simeq j_0$  \qquad
 &  \qquad $j_{min}^{\infty} = \displaystyle j_0\frac{(1+\eta)}{2b}, \qquad   j_{max} > j_{max}^{\infty}$\\    
\hline\hline
\end{tabular}
\end{table*}

\section{Characteristics}

\begin{figure}[b]
\centering
\includegraphics[width=7.5cm]{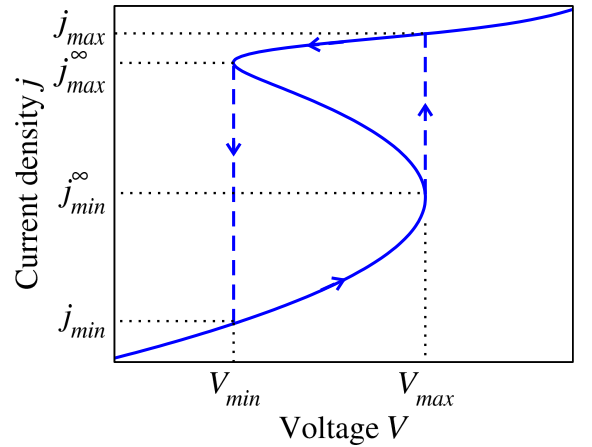}
\caption{Qualitative view of the G-FET IV with the Coulomb drag and   scheme of its bistable operation. The S-shaped IV includes three branches: the lower branch with a monotonic potential distribution $\Phi < V$, the middle branch with the potential distribution shown in Fig.~1(b), and the upper branch formed due to the inclusion of the optical phonon emission, which again  corresponds to  a monotonic potential distribution seen in Fig.~1(c).  
}
\label{F2}
\end{figure}

\subsection{General equations}
Taking into account that the leakage of the QEs from the n-region associated with the drag is compensated by the conductivity current $j_n^{<} = \sigma_n(V - \Phi)/l_n$ ({\it so that the net current density in the n-region} $j_n^{>} + j_n^{<} = j_i$),
we arrive at the following equation relating the current density $j=j_i$, potential $\Phi$, and  applied voltage $V$:

\begin{eqnarray}\label{eq13}
j= je^{- K_{ac} - K_{op}}\biggl[e^{- K_{ee}} + \frac{\Phi}{T\xi}(1 - e^{- K_{ee}})\biggr]
\nonumber\\
+ \frac{\sigma_n}{l_n}(V - \Phi).
\end{eqnarray}
Considering Eq.~(3),  Eq.~(13) can be presented as a relation between the potential $\Phi$ and the bias voltage $V$

\begin{eqnarray}\label{eq14}
\Phi\biggl(1 + \eta - e^{-K_{ee}-K_{ac}- K_{op}}\biggr)
\nonumber\\
 - \frac{\Phi^2}{T\xi}e^{-K_{ac} - K_{op}}\biggl(1 - e^{- K_{ee}}\biggr)
=\eta\,V.
\end{eqnarray}
Here $\eta = (2\pi\,\sigma_n/\kappa\,v_W)(l_i/l_n) = \sigma_nl_i/\sigma_il_n$ with $\sigma_i = \kappa\,v_W/2\pi$ [see Eq.~(3)]. 
The parameter $\eta$ is actually the ratio of the i-region resistance $r_i = l_i/\sigma_i$ and the n-region resistance $r_n = l_n/\sigma_n$ (per unit length in the direction perpendicular to the current flow): $\eta = r_n/r_i$.  
At the QE mobility $\mu_n =  10^4$~cm$^2$/s$\cdot$V, $\Sigma_n = 5\times 10^{11}$~cm$^{-2}$, and $l_i /l_n= 0.1 - 0.5$, one obtains $\eta \simeq 1.13 - 5.66$.

Equation~(14) yields the following source-drain voltage-current characteristics

\begin{eqnarray}\label{eq15}
\frac{j}{j_0}\biggl(1 + \eta - e^{-K_{ee}-K_{ac}-K_{op}}\biggr)\nonumber\\ - b\biggl(\frac{j}{j_0}\biggr)^2
e^{- K_{op}}\biggl(1 - e^{- K_{ee}}\biggr)\biggr]
=\eta\frac{V}{V_0}.
\end{eqnarray}
with $j_0 = v_W\kappa\hbar\omega_0/2\pi\,el_i$, $V_0 = \hbar\omega_0/e$, and $b = (\hbar\omega_0/T\xi)e^{-K_{ac}}\simeq 
(\hbar\omega_0/\mu)e^{-K_{ac}}$ is the Coulomb  drag parameter (see Appendix A). 
This parameter determines the ratio of the current created by the DQEs and the current of the injected BEs, which is equal to $\xi = b(j_i/j_0)e^{-K_{op}}(1 - e^{-K_{ee}})$. When the potential drop across the i-region $\Phi < V_0$, the optical phonon emission is blocked, i.e., $K_{op} = 0$, $\xi \simeq b(j_i/j_0)(1 - e^{-K_{ee}})\simeq b(j_i/j_0)$ (see below).    
The parameters $\eta$ and  $b$
 are crucial for the distribution of the potential drops across the i- and n-regions.
Due to the dependence of the parameter $b$ on the Fermi energy $\mu$, this parameter is controlled by the gate voltage $V_g$: $b \propto \mu^{-1} \propto V_g^{-1/2}$. 
Setting  $\hbar\omega_0 = 200$~meV, $\mu = 60$~meV ($\Sigma_n \simeq 6\times 10^{11}$~cm$^{-2}$), and $K_{ac} = 0.25$, we obtain $b \simeq 2.67$. At  $\kappa = 4$ and  $l_i = (0.5 - 1.0)~\mu$m, we arrive at the following estimate:
$j_0 \simeq (1.41 - 2.82)\times 10^{-4}$~A/$\mu$m.

\subsection{Low current densities} 
 
 In the voltage and current density ranges where $ \Phi < V_0$, $j < j_0$, and  $K_{op} \simeq 0$, so that the optical phonon emission is not involved in the IV formation,
 we obtain from Eqs.~(14) and (15)

\begin{eqnarray}\label{eq16}
\frac{j}{j_0}\biggl(1 + \eta - b\frac{j}{j_0}\biggr)
=\eta\frac{V}{V_0}.
\end{eqnarray}
Here and in the following we omit the term $e^{-K_{ee}} \ll 1$.

As follows from Eq.~(16), at a certain  voltage $V=V_{max}$, where 

\begin{eqnarray}\label{eq17}
V_{max} = V_0\frac{(1+\eta)^2}{4b\eta},
\end{eqnarray}
one obtains $dj/dV |_{V=V_{max}} = \infty$. This point corresponds
to $j = j_{min}^{\infty}$

\begin{eqnarray}\label{eq18}
j_{min}^{\infty} = j_0\frac{(1+\eta)}{2b}.
\end{eqnarray}
Equation~(16) describes the IV lower branch in Fig.~2. It also describes  the IV middle branch
if the latter exist, that happen if $V_{min} < V_{max}$ and $j_{min}^{\infty} < j_{max}^{\infty}$ as seen in Fig.~2.

Naturally, in the absence of the Coulomb drag ($b =0$), such a voltage point does not exist ($V_{max} \propto 1/b$ tends to infinity). Equation~(16) also corresponds to $dj/dV  < 0$, i.e., the negative differential conductivity,  when 
$j \gtrsim  j_{min}^{\infty}$.

\subsection{ High current densities} 
When $\Phi \gtrsim V_0$, $j \gtrsim j_0$, and $K_{op} \geq 0$. In this case, the optical phonon emission starts to play a substantial role. Accounting for
such an emission, from   
 Eq.~(14) we obtain the following generalization of Eq.~(16):

\begin{eqnarray}\label{eq19}
\frac{j}{j_0}(1+\eta)\nonumber\\
 - b\biggl(\frac{j}{j_0}\biggr)^2
\exp\biggl[-\overline{K}_{op}\biggl(\frac{j}{j_0}-1 +F\biggr) \Theta\biggl(\frac{j}{j_0}-1\biggr) \biggr]\nonumber\\
=\eta\frac{V}{V_0},
\end{eqnarray}
where we have introduced    the normalized electron Fermi energy $F = \mu/\hbar\omega_0$. In particular, Eq.~(19) describes the IV upper branch with $V_0 \lesssim \Phi < V$, i.e., characterized by a monotonic potential distribution shown in Fig.~1(c). 

The IV governed by Eq.~(19) exhibits the point near the threshold of the optical phonon emission, where  $V= V_{min}$, $j = j_{max}^{\infty}$, corresponding to 
$dj/dV |_{V=V_{min}} = \infty$, for relatively small $F$,  are close to

\begin{eqnarray}\label{eq20}
V_{min} \simeq V_0\frac{(1 + \eta  - b)}{\eta},
\end{eqnarray}

\begin{eqnarray}\label{eq21}
j_{max}^{\infty} = j_0.
\end{eqnarray}

If $V \gg V_0$, one can expect that $\Phi$ markedly exceeds $V_0$, so that $K_{op} \gg 1$, and the drag effect is suppressed by the relaxation of the BE momentum due to the optical phonon emission. In such a limit,
the high-voltage section of the IV becomes monotonically rising.

\subsection{IV peculiar points}
As follows from the above analysis, the IVs exhibit the following peculiar points  (for $K_{ee}\gg 1$), see also  Table I:\\

(a) $V= 0$ and $j=0$;\\

(b) $V = V_{min} = \displaystyle V_0\frac{1+\eta -b}{\eta}$ and 

$j= j_{min} = \displaystyle j_0\frac{1+\eta -b}{b}$;

(c) $V = V_{max} = \displaystyle V_0\frac{(1+\eta)^2}{4b\eta}$ and 
 $j= j_{min}^{\infty} =~\displaystyle j_0\frac{1+\eta}{2b}$,
 
 with $dj/dV|_{V = V_{max}}  = \infty$;\\

(d) $V = V_{max} = \displaystyle V_0\frac{(1+\eta)^2}{4b\eta}$  and $j=  j_{max} >j_{max}^{\infty}$;\\

(e) $V = V_{min} \simeq \displaystyle V_0\frac{(1+ \eta - b)}{\eta}$ and $j = j_{max}^{\infty} \simeq j_0$

 with $dj/dV|_{V = V_{min}}  = \infty$.\\

\begin{figure}[b]
\centering
\includegraphics[width=7.5cm]{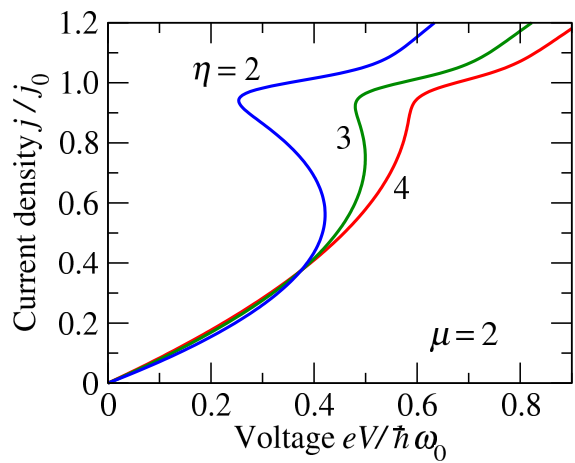}
\caption{G-FET normalized IVs ($j/j_0$ versus $eV/\hbar\omega_0$) for  
the  Fermi energy $\mu = 60$~meV  and different parameters $\eta$. 
}
\label{F3}
\end{figure}

The net current is a monotonic function of the bias voltage if   $j_{min}^{\infty} \geq j_0$, i.e., if $(1+\eta) > 2b$.
In the opposite case $j_{min}^{\infty} < j_0$, i.e., when $(1+\eta) < 2b$,       the IVs are of the S-shaped form with a region of the negative differential conductivity $dj/dV$. The latter corresponds to the voltage range $V_{min} < V < V_{max}$.

Figure~2 shows the schematic view of the G-FET S-shaped IV (analogous to those in the following Figs.~3 - 5)  at the  parameters $\eta$ and $b$
corresponding to  $(1+\eta) < 2b$ with the indicated  peculiar points corresponding to Table I. 
In 
situations when the source-drain voltage is given, the G-FET source-drain IVs can be of the S-shape.\\

 \begin{figure}[t]
\centering
\includegraphics[width=7.5cm]{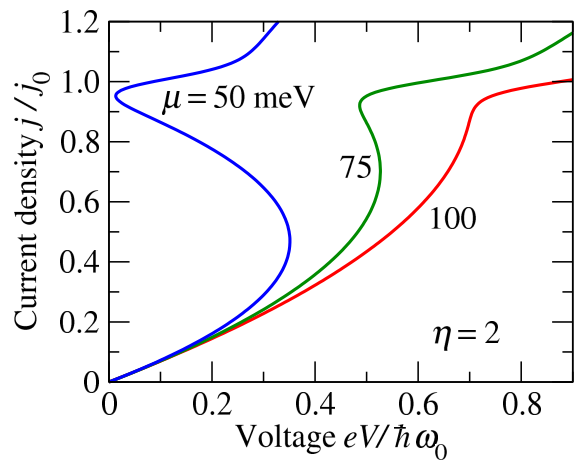}
\caption{G-FET normalized IVs ($j/j_0$ versus $eV/\hbar\omega_0$) for $\eta = 2$ and   
the Fermi energies  $\mu= 50$~meV, 75~meV, and 100~meV (the gate voltages  $V_g/d \simeq 9.2$~V/$\mu$m, 
27.7~V/$\mu$m, and  36.9~V/$\mu$m, respectively). 
}
\label{F4}
\end{figure}

\begin{figure}[t]
\centering
\includegraphics[width=7.5cm]{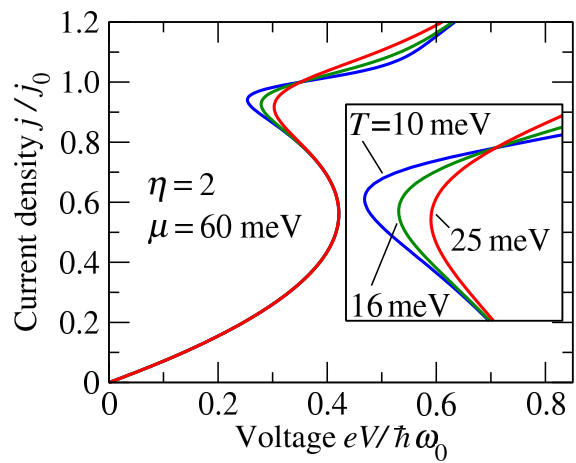}
\caption{The same characteristics as in Fig.~3, but for fixed Fermi energy $\mu = 60$~meV and different temperatures $T$.
Inset shows the IVs details near the point ($V = V_{min}$, $j = j_0$).
}
\label{F5}
\end{figure}

\subsection{Results of numerical calculations}
Figure 3 shows the IVs calculated for $K_{ee} = 5$,  $K_{op} = 0.25$, ${\cal K}_{op} = 1$,
$b = 2.67$, $F = 0.3$ ($\mu = 60$~meV), $T = 10$~meV ($\sim 115$~K)
 and different values  of other parameters  ($K_{ee}$ and $\eta$)
demonstrating their transformation from the monotonic to  S-shaped 
characteristics.                         
As seen from Fig.~3, an increase in $\eta$ (for example, due to a decrease in the n-region resistance)
leads to a weakening of the 
S-shape with a shift of $V_{min}$ toward larger values.

As follows from Eq.~(19),
the IV shape varies with changing parameter $b$, i.e., with  changing the  Fermi energy $\mu $, which, in turn, depends on   the  gate voltage  
$V_g$. The variation of $\mu$ results in
the variation of not only the parameter $b \propto \mu^{-1}\propto V_g^{-1/2}$, characterizing the drag effect, but the variation
of the parameter $F \propto \mu \propto \sqrt{V_g}$, determining the density of electron states near the threshold of the optical emission, as well. Since in the G-FETs under consideration, the dominant QE scattering mechanism is associated with the acoustic phonons (short-range scattering mechanism, which is the same as for neutral impurities and point defects), the gated n-region conductivity $\sigma_n$ and, therefore, $\eta$ can be set independent of $\mu$~\cite{39,40}. 
The change in $\mu$ and, consequently in the QE density affects $K_{ee}$. However, this  can be disregarded until
$K_{ee} \gg 1$, i.e., until the QE density is not too small. 
  
Figure~4 shows the G-FET IVs calculated using Eq.~(19)  for $\eta = 2$ and different values of the Fermi energy
$\mu$. The same other parameters and the temperature are assumed as for Fig.~3.
One can see that the IV shape is fairly sensitive to the QE Fermi energy in the gated n-region
$\mu$, i.e., depends on the QE density $\Sigma_n$ and, hence, on the gate voltage $V_g$. 
An increase in $\mu$ can result in the transformation from the S-shaped IVs to the monotonic IVs.
This is attributed to a weakening of the drag effect with increasing $\mu$ (see below). 
 
 An increase in the temperature beyond $T = 10$~meV
 leads to the IVs with a less pronounced S-shape, although such  characteristics could be obtained even at room temperature if the parameters are chosen properly, in particular, by chosing sufficiently, small $\eta$ and $\mu$ 
 ($\mu \lesssim 75$~meV). Indeed, choosing $\mu = 60$~meV and $\eta = 2$ (other parameters are the same as in Figs.~3 and 4),
 we arrive at the S-shaped IVs shown in Fig.~5, corresponding 
to the temperature range from $T = 10$~meV to $T=25$~meV. As seen, for the latter set of the parameters  the S-shape can be preserved
even at room temperature. 
The temperature smearing of the  threshold of the optical phonon emission leads to a small deviation
(for moderate values $F$) of the peculiar point positions from the values given in Table~I.
The effects of the Fermi energy and the temperature variations on the IVs are attributed to the Coulomb drag parameter $b$ versus $\mu$ and $T$ dependences. Figure~6 shows examples of these dependences calculated using Eq.~(A7) in Appendix A.
Assuming that $\tau_{ac} \propto T^{-1}$, in the calculations of $b$ we set $K_{ac} = 0.25(T [ {\rm meV}]/25)$.
One can see that a decrease  in $\mu$ and $T$ provides a rise of $b$ (and, therefore, the IVs with a more pronounced  S-shape).

\section{Current switching  by the voltage pulses}

 The S-shaped IVs with hysteresis can enable the bistable operation controlled by the source-drain voltage.
At the fixed source-drain voltage ${\overline V}$ in the voltage range $V_{min} < {\overline V} < V_{max}$, there are two branches of the stable states: the "low" stable with the current densities $0< j^{(low)} < j_{min}~{\infty}$ and the "high" stable with $j_{max}^{\infty} < j^{(high)} < \infty$. The stability of these states is due  the positive differential conductivities
$\sigma_D^{(low)} = dj^{(low)}/dV$ and $\sigma_D^{(high)} = dj^{(high)}/dV$ at the pertinent branches.
In contrast, the states in the intermediate branch $j_{min}^{\infty} < j^{(int)} < j_{max}^{\infty}$ are unstable (see below). 

The transition from the low state to the high state requires the voltage pulse $\Delta V > V_{max} - {\overline V} >0$.
The reverse transition can be realized by applying the voltage pulse $\Delta V < {\overline V} - V_{min} < 0$.
The pulse duration should be sufficiently longer than the characteristic time of the temporal relaxation of the electron system $\tau_{rc}$. This time, as  is estimated in the next section, can be an order of a few ps.

Hence, the G-FETs with the S-shaped IVs  can be used for the frequency multiplication of the incoming signals. A strong IV nonlinearity at certain applied voltages can be also used for
the signal rectification and, therefore, for the signal detection.

\begin{figure}[t]
\centering
\includegraphics[width=7.5cm]{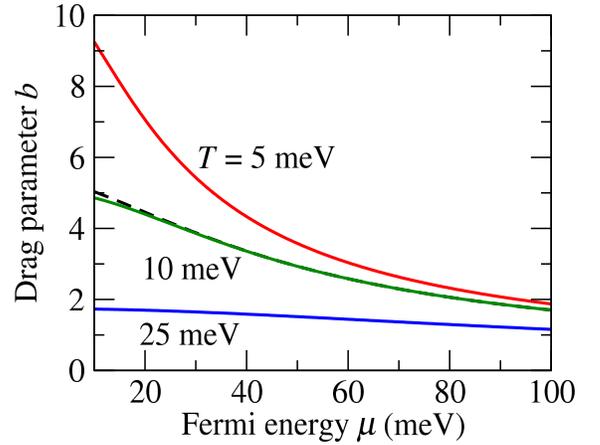}
\caption{The Coulomb drag parameter $b$ versus Fermi energy
at different temperatures. Dashed line corresponds to the dependence calculated disregarding a small effect of the quasi-equilibrium holes.}
\label{F6}
\end{figure}

\section{Aperiodic instability of uniform current flow}

In the devices with the S-shaped IVs the current tends to  filamentation under the condition when the net terminal current is fixed.
This is due to the instability of the uniform state of the electron plasma toward
the spatial perturbations in the in-plane $y$-direction, perpendicular to the current flow (in the $x$-direction).

Let us consider the dynamic behavior of the electron system.  
Introducing the normalized average current density   ${\overline J} = I/Lj_0$, where $I$ is the net current
through the G-FET (which is maintained to be fixed) chosen to be such that  ${\overline J}$ is in the range of the negative differential conductivity), $L$ is the width in the $y$-direction, and introducing

\begin{eqnarray}\label{eq22}
\tau_{rc} = \frac{c_n}{(\sigma_i/l_i + \sigma_n/l_n)}= \biggl(\frac{l_il_n}{2dv_W}\biggr)\frac{1}{(1+\eta)},
\end{eqnarray}
\begin{eqnarray}\label{eq23}
 {\cal L} = \frac{l_il_n}{2(\sigma_i/\sigma_n + l_i/l_n)} =
l_n\sqrt{\frac{\eta}{2(1+\eta)}},
\end{eqnarray}
we present an equation governing the spatio-temporal variations in the gated n-region given in Appendix B
[Eq.~(B1)]
 in the following form:

\begin{eqnarray}\label{eq24}
-\tau_{rc}\frac{\partial }{\partial t}\biggl(\frac{j}{j_0}\biggr) + {\cal L}^2\frac{\partial^2 }{\partial y^2}\biggl(\frac{j}{j_0}\biggr)= - \biggl[{\biggl(\frac{j}{j_0}\biggr) - \overline J}\biggr]\nonumber\\
 - 
\frac{b}{(1+\eta)}\biggl[\biggl(\frac{j}{j_0}\biggr)^2- {\overline J}^2\biggr].
\end{eqnarray}
The quantity $\tau_{rc}$ is a product of the gated n-region capacitance $c_n$ and the G-FET source-drain resistance $r =(\sigma_i/l_i + \sigma_n/l_n)^{-1}$.

Now we focus  on the stability of the uniform current flow with $J = {\overline J}$.
Assuming that the potential  $j = {\overline J}j_0 + \delta j e^{i(qy-\omega t)}$, where 
 $q$ and 
$\omega$ are the wave number and the  frequency of the perturbation, respectively,  
 we obtain from Eq.~(23)
the following 
 dispersion equation for the transit potential perturbations:

\begin{eqnarray}\label{eq25}
(i\omega \tau_{rc} -q^2{\cal L}^2)\delta j= \biggl(1 - \frac{2b{\overline J}}{1+ \eta}\biggr)\delta j.
\end{eqnarray}
When
\begin{eqnarray}\label{eq26}
{\overline J} > \frac{(1+\eta)}{2b},
\end{eqnarray}
the right-hand side of Eq.~ (25) is negative. This corresponds to ${\overline J}$ 
 falling to  the current range  $j_{min}^{\infty}/j_0 < {\overline J} < 1$, in which, as mentioned above,  the differential conductivity is negative. In this current  range, Eq.~(25)  for the electron plasma perturbations increment (the grows rate)
yields

\begin{eqnarray}\label{eq27}
{\rm Im}~\omega  = \frac{2b{\overline J} - 1 -\eta  - q^2{\cal L}^2 }{\tau_{rc}},
\end{eqnarray}
which is positive  for sufficiently small wave numbers $q$, i.e., for sufficiently
long perturbations.  However, the perturbation length  is limited by the device size, $L$, in the $y$-direction.

Since, according to Eq.~(25),  Re~$\omega =0$, inequality (26) corresponds to a temporal aperiodic  rise of the plasma perturbations (aperiodic plasma instability). Hence, the temporal variation of the electron system out of equilibrium, including the transformation of the current spatial distribution and the duration of the switching process is characterized by the time $\tau_{rc}$ given by Eq.~(22).
Setting, for example, $\eta = 2$, $l_i= 1~\mu$m, $l_n = 1~\mu$m, and $d= (0.05 -0.10)~\mu$m, for the characteristic time, $\tau_{rc}$, determining the time scale of the dynamic processes in the G-FETs, we obtain  $\tau_{rc} \simeq 1.67 -3.33$~ps. 

Setting ${\overline J} = 0.875$ (that corresponds to the dc potential at $x=0$ equal to  
${\overline \Phi}_0 = 175$~mV, i.e., ${\overline \Phi}_0 < V_0 = 200$~mV), $T= 10$~meV, and $\mu = 60$~meV ($b= 2.67$),
from Eq.~(27) we find that  the plasma instability in question can occur if
$\eta \leq 3.66$.

The current instability associated with the S-type IVs is akin to those
predicted by B. K. Ridley (see, for example,~\cite{41,42,43,44,45}), although it is caused by a  different
mechanism, namely, by the electron drag. 

As can be concluded from Eq. (27), the spatial scale of the current filaments is determined by the characteristic length 
${\cal L}$ given by Eq.~(23). %
Hence, the filamentation is possible when the width of the G-FET in the y-direction $L \gg {\cal L}$. 
Setting $l_n = (1- 2)~\mu$m and $\eta = 2$, we find ${\cal L} \simeq (0.57 - 1.15 )~\mu$m. 
Depending on the boundary conditions at the G-FET  edges (in the $y$-direction, $y = 0$ and $y=L$),
the arising filaments cam be either stationary or pulsating. The formation of the nonlinear filament structure might change
the source-drain voltage drop at the fixed net current. In the case of the pulsating filaments, the source-drain voltage can comprise an ac component.  

\section{Comments}

\subsection{Origin of the S-shaped IVs}

As shown above, at sufficiently strong drag effect (large $b$),
the IVs can be of the S-shape. This is associated with the following
two reasons: First, 
if the potential drop across the i-region $\Phi$ is smaller than
the voltage corresponding  to the optical phonon emission,
there the IV ambiguity with two possible values of the current: a relatively low with  a small  contribution of the DQEs and rather large with a marked contribution of the DQEs [this ambiguity is described by Eq.~(16)]. In the first case, the potential difference $V - \Phi >0$ removes the injected BEs that have accumulated in the n-region. Such a  low IV  branch corresponds to an elevated source-drain voltage $V$ [see Fig.~1(c)].
In contrast, in the second case, the DQE current through the n-region is  compensated by the reverse current injected from the drain. The latter requires $V - \Phi <0$, i.e., a lowered voltage $V$ as seen from Fig.~1(b).
Second, when $\Phi$ is sufficient for the emission of optical phonons by the injected BEs (near the point separating the i- and n-region), the drag suppressed, and the transport become normal, i.e., with  the monotonic potential distribution. The latter situation     
corresponds to the upper  branch of the S-shaped IV.
In the less probable case of too large parameter $b$, the upper branch can appear because of reflection of the DQEs by a strong  braking electric field $(\Phi - V)/l_n$.

\subsection{Electron injection and transit-time delay}
In the case of the  lateral n$^+$-contact, the BE injection is limited by the two-dimensional space-charge in the i-region. The G-FETs with the BE tunneling injection through the Schottky contact can exhibit a similar behavior. However, in the latter case, the $j_i$ versus $\Phi$ relation can be different [a nonlinear in contrast to Eq.~(3)]. This can lead to a modification of the IVs in comparison with derived above. 
 
At AC voltage, the density of the BE current injected into the n-region exhibits a delay due to the finite transit time $\tau_{tr} = l_i/v_W$ of the BEs across the i-region. Such a BE transit delay can, in principle,  affect the transient processes in the G-FETs under consideration, in particular, the dynamic of the instability considered above.

According to the Shockley-Ramo theorem~\cite{46,47},
one needs to replace the quantity $j_i = \sigma_i\Phi/l_i$ (which constitutes the quasi-stationary current density) in the right-hand side of Eq.~(9) by the current density of  the BEs  propagating across the i-region.
Therefore, the ac component of the induced current density can be presented as~\cite{6,7}

\begin{eqnarray}\label{eq28}
j_{ind} =  \frac{2}{\pi}j_i \int_0^1\frac{ds\,
\displaystyle e^{i\omega\tau_{tr}s}}{\sqrt{1-s^2}}
\simeq j_i\biggl[J_0(\omega\tau_{tr})+  \frac{2i}{\pi}\omega\tau_{tr}
\biggr], 
\end{eqnarray}
where  $J_0(s)$ is the Bessel function and
the factor $2/\pi\sqrt{1-s^2}$ under the integral appears due to the  electric field created by the BEs in the case of the "blade-shaped" highly conductive n$^+$- and gated n-regions~\cite{48}.

According to Eq.~(28), the relative role of the  transit-time effect is weak in comparison with the effect of the gated n-region RC-recharging is characterized by the ratio $2\tau_{tr}/\pi\tau_{rc}$. Taking into account Eq.~(22),
we find $2\tau_{tr}/\pi\tau_{rc}  = 4(1+\eta)d/\pi\,l_n$. 

For $\eta= 2$, $l_n = 1~\mu$m, $d = 0.05 -0.10~\mu$m, one obtains  $2\tau_{tr}/\pi\tau_{rc} \simeq 0.19 - 0.38  < 1$.
Since the latter inequality is 
normally satisfied for the G-FETs with realistic parameters, we disregarded the BE transit delay,
although this effect can lead to a moderate decrease of the instability increment. 

\subsection{Plasmonic resonance effects}
The two-dimensional electron system in the gated n-region of the GL channel can exhibit the plasmonic resonances corresponding to the plasma oscillation frequencies $\Omega \propto d^{1/4}V_g^{1/4}/l_n$ and its harmonics~\cite{49}. The excitation of the plasma oscillations is possible when the source-drain voltage $V$ comprises the ac component with the frequency $\omega \simeq \Omega$.
This component can be associated with the incident radiation received by an antenna. This effect combined with the pronounced IV nonlinearity should lead to a resonantly large rectified current, which can be used for the detection of the incoming radiation.
According to the estimate of the plasma frequency $\Omega$, it can be in the terahertz frequency (THz) range.
Due to the positive feedback between the  currents injected to the i-region from the source and the reverse current injected to the n-region from the drain, one might expect the plasma instability
of the net steady-state source-drain current resulting in the self-excitation of the THz plasma oscillations~\cite{50}. 
 However, the analysis of such effects is beyond the scope of the present paper.   	

\subsection{Technological aspects}

The crystallographic quality of graphene synthesized by a popular engineering method of the thermal decomposition from the SiC substrate is now approaching the high end of those for exfoliated graphene~\cite{51}, in particular, exhibiting the BE transport [10, 11]. The G-FET (similar to that under consideration in this paper) process technology is getting matured for both semiconductor integrated device processes based on e-beam lithography and gate stack formation with the plasma chemical vapor deposition (CVD)~\cite{52} or the atomic layer deposition (ALD) and exfoliation and dry-transfer in hBN/graphene/hBN van der Waals hetero-stacking for the gate stack~\cite{1}. The processed GL channels in those G-FET devices with sub-micrometer dimensions exhibit field-effect mobilities beyond 100,000 cm2/Vs~\cite{51,52}.

\section{Conclusions}

We proposed and evaluated the characteristics of a lateral n$^+$-i-n-n$^+$ G-FET with the ballistic transport of the electrons injected from the source n$^+$-region into the i-region. We demonstrated that  the ballistic electrons entering the n-region can effectively drag the quasi-equilibrium electrons toward the drain if the electron-electron scattering in the gated n-region prevails over the impurity and acoustic phonon scattering.
The  Coulomb ballistic-equilibrium electron drag in question 
with the electron current multiplication
is  associated with the linearity of the electron energy dispersion law in graphene.
The drag effect can result in non monotonous potential distributions in the G-FET channel and the strongly nonlinear S-type source-drain IVs. The S-type IVs might lead to the filamentation of the current
in the G-FET channel (with the stationary or pulsating filaments) and to the hysteresis phenomena, enabling the switching
between different current states. Apart from this application,
the plasmonic phenomena in the G-FETs under consideration can be used for the THz radiation detection, generation, and   signal frequency-multiplication. The latter applications require  a separate consideration.

\section*{Acknowledgments}
The work at RIEC and UoA was supported by Japan Society for Promotion of Science, KAKENHI Grant Nos. 21H04546, 20K20349, 
Japan; RIEC Nation-Wide Collaborative research Project No. H31/A01, Japan;
The work at RPI was supported by Office of Naval Research (N000141712976, Project Monitor Dr. Paul Maki). The authors  are grateful to D. Svintsov for very useful discussions. One of the authors (V.R.) is also thankful to Yu. G. Gurevich for helpful information. 


\section*{Appendix A. QE Coulomb drag parameter}
\setcounter{equation}{0}
\renewcommand{\theequation} {A\arabic{equation}}

The average momentum transferring from BEs to QEs (per one QE) can be presented as:

\begin{eqnarray}\label{eqA1}
<p_x> = 
\frac{j_i\Phi}{v_W^2\Sigma_n}e^{-K_{ac}-K_{op}}(1 - e^{- K_{ee}}).
\end{eqnarray}

The QE distribution function

\begin{eqnarray}\label{eqA2}
f = \biggl[1 + \exp\biggl(\frac{v_Wp + \mu - p_x <v_x>}{T}\biggr) \biggr]^{-1},
\end{eqnarray}
whee $<v_x>$ is the average drift velocity obtained by QEs due to the collisions with the BE flux, $T$
is temperature and $\mu$ is the electron Fermi energy: $\mu \simeq \hbar\,v_W\sqrt{\kappa\,V_g/4ed}$.
The latter yields the relation between $<p_x>$ and $<v_x>$:

\begin{eqnarray}\label{eqA3}
<p_x> = \frac{\displaystyle\int \frac{ dp_ydp_x p_x}{\biggl[1 + \displaystyle\exp\biggl(\frac{v_Wp + \mu - p_x <v_x>}{T}\biggr) \biggr]}
}{\displaystyle
\int \frac{dp_xdp_y}{\biggl[1 + \displaystyle\exp\biggl(\frac{v_Wp + \mu}{T}\biggr) \biggr]}}\nonumber\\
 \simeq \frac{<v_x>T}{v_W^2}\xi(\mu/T).
\end{eqnarray}
where 
\begin{eqnarray}\label{eqA4}
\xi(\mu/T) 
=\displaystyle\frac{3}{2}\frac{{\cal F}_2(\mu/T)}{{\cal F}_1(\mu/T)}.
\end{eqnarray}
Here ${\cal F}_n(\eta) = \int_0^{\infty}duu^n[1 + \exp(u-\eta)]^{-1}$ is the Fermi-Dirac integral.
At $\mu/T \gg 1$, $\xi(\mu/T) \simeq \mu/T$. 

Hence,
\begin{eqnarray}\label{eqA5}
j_n^{>} =  e\Sigma_n<v_x> \simeq  j_ie^{-K} 
\nonumber\\
+ \frac{j_i\Phi}{T\xi}e^{-K_{ac}-K_{op}}(1 - e^{- K_{ee}}).
\end{eqnarray}
Using Eq.~(A4), the quantity $b = (\hbar\omega_0/T\xi)e^{-K_{ac}}(1 - e^{- K_{ee}}) \simeq (\hbar\omega_0/T\xi)e^{-K_{ac}}$, 
which we call as the Coulomb drag parameter, 
is presented as
  
\begin{eqnarray}\label{eqA6}
b = \frac{3\hbar\omega_0}{2T}e^{-K_{ac}}\frac{{\cal F}_1(\mu/T)}{{\cal F}_2(\mu/T)}.
\end{eqnarray}
When $\mu > T$, $b \simeq (\hbar\omega_0/\mu)e^{-K_{ac}} \simeq (\hbar\omega_0/\mu)$.
Since Fermi energy $\mu$ depends on the QE density $\Sigma_n$, $\mu$ and, therefore, $b$ are controlled by the gate voltage $V_g$.

If the value of $\mu$ approaches to the Dirac point, the drag of the quasi-equilibrium holes (QHs) 
can become crucial. 
This is because the QHs are dragged by the BEs to the same direction partially neutralizing the current of the dragged QEs.

Considering that the QH Fermi energy is equal to $-\mu$, one can obtain the expression for the drag  parameter $b$ replacing Eq.~(A6):

\begin{eqnarray}\label{eqA7}
b =\frac{2\hbar\omega_0}{3T}e^{-K_{ac}}\biggl[ \frac{{\cal F}^2_1(\mu/T)}{{\cal F}_2(\mu/T)}  -\frac{{\cal F}^2_1(-\mu/T)}{{\cal F}_2(-\mu/T)}\biggr]\nonumber\\
\times[{\cal F}_1(\mu/T)+ {\cal F}_1(-\mu/T)]^{-1}.
\end{eqnarray}
Equation~(A7) does not account for the mutual electron-hole drag~\cite{18}. The latter should lead to a somewhat smaller value of $\xi(\mu/T)$
in comparison with Eq.~(A6).
Thus the QHs weaken  the drag current multiplication. Although for the G-FETs with the parameters used in the main text,
this is negligible.

For different devices with the two-dimensional carriers but with the quadratic dispersion
(having the 2D channels in the heterostructures made of the standard materials and the graphene bilayer heterostructures), $<v_z> = <p_z>/m$, where $m$ is the carrier effective mass) and $j_n^{>} =j_i$,
so that there is no electron current multiplication.

\section*{Appendix B. Spatio-temporal variations of electron system in the gated region}
\setcounter{equation}{0}
\renewcommand{\theequation} {B\arabic{equation}}

The electron charge  in the diode active region ($0 < x < l_n$) $Q = -c_n\Phi$,
where $c_n = \kappa\,l_n/4\pi\, d$ is the gated n-region capacitance,
obeys the following equation:

\begin{eqnarray}\label{eqB1}
\frac{\partial Q}{\partial t} + \frac{\sigma_nl_n}{2}\frac{\partial^2 \Phi}{\partial y^2}= \frac{\sigma_i\Phi}{l_i}\biggl(1 - \frac{e\Phi}{T\xi}e^{-K_{ac} -K_{op}}\biggr)\nonumber\\ 
- \frac{\sigma_n}{l_n} (V - \Phi).
\end{eqnarray}
The factor 1/2 in the second term in the left-hand side of Eq.~(B1), appears
because the potential in the n-region varies between  $\varphi|_{x=0} = \Phi$
and $\varphi|_{x=l_n} = V$ (approximately linearly).
Equation~(B1) can be presented as

\begin{eqnarray}\label{eqB2}
\frac{c_nl_i}{\sigma_i}\frac{\partial}{\partial t}\biggl(\frac{j}{j_0}\biggr) + \frac{\sigma_n}{\sigma_i}\frac{l_nl_i}{2}\frac{\partial^2 }{\partial y^2}\biggl(\frac{j}{j_0}\biggr)
=\frac{j}{j_0}(1+ \eta)\nonumber\\
 -b\biggl(\frac{j}{j_0}\biggr)^2\exp\biggl[-\overline{K}_{op}\biggl(\frac{j}{j_0}-1 +F\biggr) \Theta\biggl(\frac{j}{j_0}-1\biggr) \biggr]\nonumber\\
-\eta\frac{V}{V_0}.
\end{eqnarray}

In the case of the steady-state uniform current flow with $j < j_0$ when $K_{op} =0$, Eq.~(B1) is reduced to Eq.~(24). 
If the average  current density through the G-FET  $I/L$ and its normalized value $I/Lj_0$ are given, Eq.~(B2) can also be presented in the following form:

\begin{eqnarray}\label{eqB3}
-\tau_{rc}
\frac{\partial}{\partial t}\biggl(\frac{j}{j_0}\biggr) 
+{\cal L}^2\frac{\partial^2 }{\partial y^2}\biggl(\frac{j}{j_0}\biggr)
\nonumber\\
=\biggl(\frac{j}{j_0}\biggr) - \frac{b}{(1+\eta)}\biggl[\biggl(\frac{j}{j_0}\biggr)^2 - {\overline J}^2\biggr].
\end{eqnarray}
Here $\tau_{rc}= c_n/(\sigma_i/l_i+\sigma_n/l_n)$ and ${\cal L}^2 =l_il_n/2(\sigma_i/\sigma_n + l_i/l_n)$.


\begin{thebibliography}{1}

\bibitem{1}
S. Boubanga-Tombet, W. Knap, D. Yadav, A. Satou, D. B. But, V. V. Popov, I. V. Gorbenko, V. Kachorovskii, and T. Otsuji,
\lq\lq Room temperature amplification of terahertz radiation by grating-gate graphene structures,\rq\rq
 Phys. Rev. X {\bf 10},  031004-1-19 (2020).
 
 \bibitem{2}
 J. A. Delgado-Notario, V. Clericò, E. Diez, J. E. Velazquez-Perez, T. Taniguchi, K. Watanabe, T. Otsuji, and Y. M. Meziani,
\lq\lq Asymmetric dual grating gates graphene FET for detection of terahertz radiations,\rq\rq
APL Photon. {\bf 5}, 066102-1-8 (2020).
 
\bibitem{3} 
V. Ryzhii, T. Otsuji, and M. S. Shur,
\lq\lq  Graphene based plasma-wave devices for terahertz applications,\rq\rq
Appl. Phys. Lett.{\bf 116}, 140501-1-6 (2020).
 
 \bibitem{4}
 M. Ryzhii and V. Ryzhii,
 \lq\lq 
Injection and population inversion in electrically induced p–n junction in graphene with split gates, \rq\rq
 Jpn. J. Appl. Phys. {\bf 46}, L151 (2007).

\bibitem{5}
M. Ryzhii, V. Ryzhii, T. Otsuji, V. Mitin, and  M. S. Shur,
\lq\lq Electrically induced n-i-p junctions in multiple graphene layer structures,\rq\rq
Phys. Rev. B {\bf 82}, 075419 (2010).



\bibitem{6}
 V. Ryzhii, M. Ryzhii, V. Mitin, and M. S. Shur,
\lq\lq Graphene tunnelinhg transit-time terahertz oscillator based on electrically induced p-i-n junctions,\rq\rq
Appl. Phys. Express
 {\bf 2}, 034503 (2009).

 \bibitem{7} V. L. Semenenko, V. G. Leiman, A. V. Arsening, V. Mitin, M. Ryzhii, T. Otsuji, and V. Ryzhii,
\lq\lq Effect of self-consistent electric field on characteristics of graphene p-i-n tunneling transit-time diodes,\rq\rq
J. Appl. Phys. {\bf 113}, 024503 (2013).

\bibitem{8}
M. S. Shur and L. F. Eastman, 
\lq\lq Ballistic transport in semiconductor at low temperatures for low-power high-speed logic,\rq\rq
IEEE Trans. Electron Devices {\bf 26}, 1677-1683 (1979).


\bibitem{9}
V. I. Ryzhii, N. A. Bannov, and V. A. Fedirko, \lq\lq Ballistic and quasi ballistic transport in semiconductor structures (review)\rq\rq, Sov. Phys. Semicond. {\bf 18}, 481 (1984).
%





\bibitem{10}
A. S. Mayorov, R. V. Gorbachev, S. V. Morozov, L. Britnell, R. Jalil, 
L. A. Ponomarenko, P. Blake, K. S. Novoselov, K. Watanabe, T. Taniguchi,
and A. K. Geim,
\lq\lq Micrometer-scale ballistic transport in encapsulated graphene
at room temperature,\rq\rq
Nano Lett. {\bf 11}, 2396-2399 (2011).


\bibitem{11}
L. Banszerus, M. Schmitz, S. Engels, M. Goldsche, K. Watanabe, T. Taniguchi, B. Beschoten, and C. Stampfer,
\lq\lq Ballistic transport exceeding 28 $\mu$m in CVD grown graphene,\rq\rq
Nano Lett. {\bf 16}, 1387-1391 (2016).



\bibitem{12} 
A. A. Andronov and V. A. Kozlov,
\lq\lq  Low-temperature negative differential microwave conductivity in semiconductors following elastic scattering of electrons,\rq\rq
JETP Lett. {\bf 17}, 87 (1973).

\bibitem{13} 
V. L. Kustov, V. I. Ryzhii, and Yu. S. Sigov, \lq\lq Nonlinear plasma instabilities in semiconductors subjected to strong electric fields in the case of inelastic scattering of electrons by optical phonons,\rq\rq 
Sov. Phys. JEPT {\bf 99} (1980).

\bibitem{14} 
L. E. Vorob’ev, S. N. Danilov, V. N. Tulupenko, and  D. A. Firsov,
\lq\lq Generation of millimeter radiation due to electric-field-induced electron-transit-time resonance in indium phosphide,\rq\rq
JEPT Lett. {\bf 73}, 219 (2001).
 
\bibitem{15} 
S. Sekwao and J. P. Leburton,
\lq\lq 
Terahertz harmonic generation in graphene,\rq\rq
Appl. Phys. Lett. {\bf 106}, 063109 (2015). 

\bibitem{16} 
 A. A. Andronov and V. I. Pozdniakova,
\lq\lq
Terahertz dispersion and amplification under electron streaming in graphene at 300 K, 
\rq\rq
Semiconductors {\bf 54}, 1078-1085 (2020).

%


\bibitem{17}
X. Li, E. A. Barry, J. M. Zavada, M. Buongiorno Nardelli, and K. W. Kim,
\lq\lq Influence of electron-electron scattering on transport characteristics in monolayer graphene,\rq\rq
Appl. Phys. Lett. {\bf 97}, 082101 (2010).

\bibitem{18}
D. Svintsov, V. Vyurkov, S. Yurchenko,  T. Otsuji, and V. Ryzhii, 
\lq\lq Hydrodynamic model for electron-hole plasma in graphene,\rq\rq
J. Appl. Phys. {\bf 111}, 083715 (2012).




\bibitem{19}
D. Svintsov, V. Ryzhii, A. Satou, T. Otsuji, and V. Vyurkov,
\lq\lq Carrier-carrier scattering and negative
dynamic conductivity in pumped
graphene,\rq\rq
Opt. Express {\bf 22}, 19873  (2014).

\bibitem{20}
T. J. Gramila, J. P. Eisenstein, A. H. MacDonald, L. N. Pfeiffer, and K. W. West, \lq\lq Mutual friction between parallel two-dimensional electron systems,\rq\rq Phys. Rev. Lett. {\bf 66}, 1216 (1991).

\bibitem{21}
U. Sivan, P. M. Solomon, and H. Shtrikman, \lq\lq Coupled electron-hole transport,\rq\rq Phys. Rev. Lett. {\bf 68}, 
1196 (1992).

\bibitem{22}
M. Schütt, P. M. Ostrovsky, M. Titov, I. V. Gornyi, B. N. Narozhny, and A. D. Mirlin,
\lq\lq Coulomb drag in graphene near the Dirac point,\rq\rq Phys. Rev. Lett. {\bf 110}, 026601 (2013).

\bibitem{23}
R. V. Gorbachev, A. K. Geim, M. I. Katsnelson, K. S. Novoselov, T. Tudorovskiy, I. V. Grigorieva, A. H. MacDonald, S. V. Morozov, K. Watanabe, T. Taniguchi, and L. A. Ponomarenko, 
\lq\lq Strong Coulomb drag and broken symmetry in double-layer graphene,\rq\rq
Nat. Phys. {\bf 8}, 896 (2012).




\bibitem{24}
J. C. Song, D. A. Abanin, and L. S. Levitov,
\lq\lq Coulomb drag mechanisms in graphene,\rq\rq Nano Lett. {\bf 13}, 3631 (2013).

\bibitem{25}
D. Svintsov, V. Vyurkov, V. Ryzhii, and T. Otsuji,
\lq\lq Hydrodynamic electron transport and nonlinear waves in graphene, \rq\rq
Phys. Rev. B {\bf 88}, 245444 (2013).

\bibitem{26}
S. H. Abendinpour, G. Vignale, A. Principi, M. Polini, W.-K. Tse, and A. H. MacDonald,
\lq\lq Drude weight, plasmon dispersion, and ac conductivity in doped graphene sheets,\rq\rq Phys. Rev. B {\bf 84}, 045429 (2011).



\bibitem{27}
J. Li, T. Taniguchi, K. Watanabe, J. Hone, A. Levchenko, and C. R. Dean, 
\lq\lq Negative Coulomb drag in double bilayer graphene,\rq\rq Phys. Rev. Lett. {\bf 117}, 046802 (2016).

\bibitem{28}
Y. Nam, D. K. Ki, D. Soler-Delgado, and A. F. Morpurgo, \lq\lq  Electron-hole collision limited transport in charge-neutral bilayer graphene,\rq\rq
Nat. Phys. {\bf 13}, 1207 (2017).





\bibitem{29}
 D. Svintsov, \lq\lq Fate of an electron beam in graphene: Coulomb relaxation or plasma instability,\rq\rq
 Phys. Rev. B {\bf 101}, 235440 (2020).


\bibitem{30}
A. Grinberg, S. Luryi, M. Pinto, and N. Schryer, \lq\lq Space-charge-limited current in a film,\rq\rq IEEE
Trans. Electron Devices {\bf 36}, 1162 (1989).

\bibitem{31} 
S. G. Petrosyan and A. Ya. Shik, \lq\lq 
Contact phenomena in low-dimensional electron systems,\rq\rq
Sov. Phys. JETP {\bf 69}, 1261 (1989).




\bibitem{32} 
 B. Gelmont, M. Shur, and C. Moglestue, \lq\lq Theory of junction between two-dimensional electron gas and p-type  semiconductor,\rq\rq
IEEE Trans.  Electron Devices  {\bf 39}, 1216 (1992).

\bibitem{33}
D. B. Chklovskii, B. I. Shklovskii, and L. I. Glazman,
\lq\lq Electrostatics of edge channels,\rq\rq Phys. Rev. B {\bf 46}, 4026~(1992).


\bibitem{34}
M. V. Beznogov and  R. A. Suris, \lq\lq
Theory of space charge limited ballistic currents
in nanostructures of different dimensionalities,\rq\rq Semiconductors {\bf 47},  514 (2013).

\bibitem{35}
R. S. Shishir  and D. K. Ferry,
\lq\lq Intrinsic mobility in graphene,\rq\rq  J. Phys.: Cond. Mat. {\bf 21},
344201 (2009).


\bibitem{36}
E. H. Hwangand and S. Das Sarma, \lq\lq Acoustic phonon scattering limited carrier mobility in two-dimensional extrinsic graphene,\rq\rq Phys.
Rev. B {\bf 77}, 115449 (2008).

\bibitem{37}
K. M. Borysenko, J. T. Mullen, E. A. Barry, S. Paul, Y. G. Semenov, J. M. Zavada, M. B. Nardelli, and K. W. Kim,
\lq\lq First-principles analysis of electron-phonon interactions in graphene,\rq\rq
Phys. Rev. B {\bf 81}, 121412(R) (2010).

\bibitem{38}
M. V Fischetti, J. Kim, S. Narayanan, Zh.-Y. Ong, C. Sachs, D. K. Ferry, and S. J. Aboud,
\lq\lq Pseudopotential-based studies of electron transport in graphene and graphene nanoribbons,\rq\rq
J. Phys: Cond. Mat. {\bf 25}, 473202 (2013).

\bibitem{39}
F. T. Vasko and V. I. Ryzhii,
\lq\lq Voltage and temperature dependences of conductivity in gated graphene heterostructures,\rq\rq
Phys. Rev. B {\bf 76}, 233404 (2007). 

\bibitem{40}
V. Ryzhii, D. S. Ponomarev, M. Ryzhii, V. Mitin, M. S. Shur, and T. Otsuji,
\lq\lq Negative and positive terahertz and infrared photoconductivity in uncooled graphene,\rq\rq
Opt. Mat. Express {\bf 9}, 585 (2019).



\bibitem{41}
B. K. Ridley,
\lq\lq Specific negative resistance in solids,\rq\rq
Proc. Phys. Soc. {\bf 82}, 954 (1963).

\bibitem{42}
A. Blicher, {\it Field-Effect and Bipolar Power Transistor Physics}, (Ac. Press, New York, 1981).

\bibitem{43}
A. F. Volkov and Sh. M. Kogan,
\lq\lq  Nonuniform current distribution in semiconductors with negative differential conductivity,\rq\rq
Sov. Phys. JETP {\bf 25}, 1095 (1967).

\bibitem{44}
F. G. Bass, V. S. Bochkov, and  Yu. G. Gurevich, 
\lq\lq Influence of sample size
on the current-voltage characteristic in media with an ambiguous dependence
of electron temperature on field strength,\rq\rq Sov. Phys. JETP {\bf 31},
972 (1970).

\bibitem{45}
F. G. Bass, Yu. G. Gurevich, S. A. Kostylev, and N. A. Terent'eva,
\lq\lq Dynamics of electrical instabilities in a  medium with an S-type negative differential conductance,\rq\rq
Sov. Phys. Semicond. {\bf 17}, 808 (1983). 



\bibitem{46}
W. Shockley, \lq\lq Currents to conductors induced by a moving point charge,\rq\rq
J. Appl. Phys. {\bf 9}, 635 (1938).



\bibitem{47}
S. Ramo, \lq\lq Currents induced by electron motion,\rq\rq
Proc. IRE {\bf 27}, 584 (1939).

\bibitem{48}
V. Ryzhii and G. Khrenov, \lq\lq High-frequency operation of lateral hot-electron transistor,\rq\rq
IEEE Trans. Electron Devices {\bf 42}, 166 (1995).

\bibitem{49}
V. Ryzhii, A. Satou, and T. Otsuji,
\lq\lq Plasma waves in two-dimensional electron-hole system in gated graphene heterostructures,\rq\rq
J. Appl. Phys. {\bf 101}, 024509 (2007).

\bibitem{50}
V. Ryzhii, A. Satou, I. Khmyrova, M. Ryzhii, T. Otsuji, V. Mitin, and M. S. Shur,
\lq\lq Plasma effects in lateral Schottky junction tunneling transit-time terahertz oscillator,\rq\rq
J. Phys.: Conf. Ser. {\bf 38}, 228 (2006). 

\bibitem{51}
T. Someya, H. Fukidome, H. Watanabe, T. Yamamoto, M. Okada, H. Suzuki,
Yu Ogawa, T. Iimori, N. Ishii, T. Kanai, K. Tashima, B. Feng, S. Yamamoto,
J. Itatani, F. Komori, K. Okazaki, Sh. Shin, and I. Matsuda,
\lq\lq Suppression of supercollision carrier cooling in high mobility graphene on SiC(0001), \rq\rq
Phys. Rev. B {\bf 95}, 165303 (2017).



\bibitem{52}
D. Yadav, G. Tamamushi, T. Watanabe, J. Mitsushio, Y. Tobah, K.
Sugawara, A. A. Dubinov, A. Satou, M. Ryzhii, V. Ryzhii, and 
T. Otsuji,
\lq\lq Terahertz light-emitting graphene-channel
transistor toward single-mode lasing,\rq\rq
Nanophotonics {\bf 7}, 741 (2018). 7(4)




\end{thebibliography}
\end{document}